# Grand Challenges in Measuring and Characterizing Scholarly Impact

**Chaomei Chen**

College of Computing and Informatics, Drexel University, Philadelphia, PA USA

chaomei.chen@drexel.edu

## 1    Introduction

Scientists, policy makers, and the general public need to access, understand, and communicate scientific knowledge. As *Heilmeier's Catechism* advocated, researchers should be able to communicate the value of their research to the public regardless whether it is a mission to Mars or a search for a cure for cancer.

The constantly growing body of scholarly knowledge of science, technology, and humanities is an asset of the mankind. While new discoveries expand the existing knowledge, they may simultaneously render some of it obsolete. It is crucial for scientists and other stakeholders to keep their knowledge up to date. Policy makers, decision makers, and the general public also need an efficient communication of scientific knowledge.

*Scholarly Metrics and Analytics* aims to provide an open forum to address a diverse range of issues concerning the creation, adaptation, and diffusion of scholarly knowledge, and advance quantitative and qualitative approaches to the study of scholarly knowledge. The following grand challenges illustrate some of the major issues concerning the interdisciplinary community.

## 2    Grand Challenge 1: Accessibility

Scientific literature is increasingly volatile. *PLoS One* alone published 30,000 articles in 2014, an average of 85 articles per day[1]. The Web of Science has accumulated over 1 billion cited references[2]. The scale of retraction has stepped up – in one incidence publishers retracted 120 gibberish papers simultaneously (Noorden 2014). While it is easy to locate a paper that we are looking for, keeping abreast of the advances of scholarly work is a constant challenge.

In addition to the common focus on documents, more efficient and incrementally maintainable approaches should enable researchers to recognize and match information of interest beyond the constraints of the form or the language. The appropriate scope of a subject should be naturally and automatically expanded to attract documents through multiple types of intellectual linkage, such as semantic, linguistic, social, citation and usage, just as an experienced expert would do to grow his/her own oeuvre of domain expertise. In addition, the self-organized and updated oeuvre of knowledge should help us understand the significance of research at the same level of clarity as *Heilmeier's*

---

[1] http://blogs.plos.org/everyone/2014/01/06/thanking-peer-reviewers/

[2] http://stateofinnovation.thomsonreuters.com/web-of-science-1-billion-cited-references-and-counting

*Catechism*. It will be fundamentally valuable to researchers and decision makers if new techniques can help us identify the state of the art of a topic more efficiently and effectively. For example, a reader can choose any topic of interest and an intelligent system can generate a systematic review of the topic as if a panel of domain experts would produce.

## 3      Grand Challenge 2: Clarity on Uncertainty

Scientific knowledge is never free of uncertainty. It is difficult to communicate uncertainty clearly, especially on issues with widespread concerns, such as climate change (Heffernan 2007) and Ebola (Johnson and Slovic 2015). The way in which the uncertainty of scientific knowledge is communicated to the public can influence the perceived level of risk and the trust (Johnson and Slovic 2015).

A good understanding of the underlying landscape of uncertainty is essential, especially in areas where information is incomplete, contradictory, or completely missing. For instance, there is no information on how long Ebola virus can survive in the water environment (Bibby, Casson et al. 2015). If surrogates with similar physiological characteristics can be found, then any knowledge of such surrogates would be valuable. Currently, finding such surrogates in the literature presents a real challenge (Bibby, Casson et al. 2015).

Another form of uncertainty rises when new inputs alter the existing structure of scholarly knowledge. A new discovery may strengthen a previously weak or missing link as well as undermine or eliminate previously strong dependencies. Distortions may be introduced by citations and reinterpretations (Greenberg 2009) or false claims made by retracted studies (Chen, Hu et al. 2013). In many areas, damages may remain unnoticed for a long time due to the lack of efficient and systematic mechanisms.

Active researchers are aware of such uncertainties in their areas of expertise. They choose words carefully and use hedging and other rhetorical mechanisms to convey their findings in the context of uncertainty. These common practices in scholarly communication have further increased the complexity of understanding science, especially for those without relevant expertise and for computational approaches. Future developments should enable stakeholders to access scholarly knowledge with a great degree of clarity on uncertainty as well as the knowledge itself.

## 4      Grand Challenge 3: Connecting Diverse Perspectives

The vast body of scholarly knowledge is a gold mine for making new discoveries. Pioneering efforts in literature-based discovery have demonstrated the value of connecting disparate bodies of knowledge discovery (Swanson 1986, Smalheiser and Swanson 1994, Cameron, Bodenreider et al. 2013). The idea of a recombinant search in technology landscapes has a great impact (Fleming and Sorenson 2001). An array of attempts have been made more recently to enhance the process of scientific discovery with the publicly available knowledge, including detecting potentially transformative ideas and emerging trends based on structural variations (Chen 2012), atypical combinations (Uzzi, Mukherjee et al. 2013), diversity in interdisciplinary research (Rafols and Meyer 2010), systematically generating and representing hypotheses (Soldatova and Rzhetsky 2011, Malhotra, Younesi et al. 2013), and the role of analogy in connecting different scientific domains (Small 2010).

Research reveals that influential ideas share a fundamental property – they tend to be richly interlinked with other ideas (Goldschmidt and Tatsa 2005). A profound theme shared by many of the attempts is the role of divergent thinking in scientific discovery, decision making, and creative problem solving, including the assessment of research excellence and impact. The value of reconciling multiple perspectives has been long recognized and advocated (Linstone 1981). The point is not so much to enlist multiple perspectives in an interdisciplinary research team; rather, the key is to expose conflicting views on the same issue and resolve seemingly contradict evidence at a new level (Chen 2014).

To meet this challenge, new computational and analytic tools should enable researchers and evaluators to work with multiple perspectives directly. The unit of operation and analysis should focus on perspectives and paradigms as well as their premises, evidence, and chains of reasoning.

## 5 Grand Challenge 4: Benchmarks and Gold Standards

Repositories of well-documented exemplar cases analyzed from multiple perspectives should be created, maintained, and shared with the research community so as to enable researchers to test and calibrate their metrics and analytic tools as well as reflect on lessons learned from these cases. Such repositories should include the most representative examples of high-impact scientific breakthroughs, the most complex cases of retracted studies, and the most extensive scientific debates in the history of science so that researchers can reproduce findings of previous studies. In particular, original datasets or queries that generate such datasets, metadata at various levels of granularity, narratives, and analytic procedures that have been applied by various studies should be preserved and made accessible. As shared resources, they will be valuable for the development and evaluation of new metrics and analytic capabilities as well as for preserving the provenance of scientific discoveries.

The role of readily available benchmarks and gold standards is crucial for a wide variety of scholarly activities. For example, Swanson's pioneering study of the possible linkage between fish oil and Raynaud's syndrome has become an exemplar case in literature-based discovery. Many subsequent studies validate newly introduced techniques with reference to the classic case. However, despite the fact that it is widely known as a classic case in literature-based discovery, the lack of essential benchmarks and gold standards makes it difficult to perform a systematic and comprehensive validation of scholarly metrics and analytic paths without spending a considerable amount of time and effort on reconstructing the vehicle for evaluation.

We can envisage how a shared repository would enable researchers to check-out a snapshot of scientific knowledge exactly as what was available to Swanson when he conducted his classic study. The snapshot would include all the information that Swanson had used in his study and discoveries he made in his original study. In addition, the repository should register and preserve similar snapshots associated with subsequent studies inspired by Swanson's original work. Although subsequent studies may introduce new sources of data, different types of information, or a wider range of levels of granularity in comparison with previous studies, gold standards should provide a consistent framework of reference such that one can systematically assess the efficiency and effectiveness of the application of a new approach to the same problem.

As research in scholarly metrics and analytics advances, we can expect that new approaches will be able to reach scientific knowledge with a greater degree of depth and breadth than before. Consequently, the evaluation of new metrics and techniques requires gold standards at comparable levels of granularity. For instance, the novelty of a hypothesis can be established at different levels of abstraction, ranging from a simple link derived from co-occurrences of keywords, a semantic path that connect two concepts separated by many other concepts, to an even broader context that, for instance, contain information reachable with a *k*-degree of separation. These levels of detail should be made readily accessible as part of the shared benchmark and gold-standard repository.

## 6    Grand Challenge 5: Integrating Scholarly Metrics and Analytics

Scholarly metrics and qualitative studies of scientific discoveries and long-range foresights need to work together. The value of experts' opinions has been widely recognized. The challenge is in soliciting and synthesizing a wide variety of views from a diverse range of experts (Linstone and Turoff 1975, Cozzens, Gatchair et al. 2010, Linstone and Turoff 2011). As strongly advocated in the Leiden manifesto, scholarly metrics should serve the supporting role to qualitative and in-depth analytics of scholarly content and activities (Hicks, Wouters et al. 2015).

Numerous scholarly metrics have been proposed, ranging from the widely known *h*-index, citation counts with or without field normalization, to altmetrics. Scholarly metrics are meant to be universal, quantifiable, field invariant, and easy to communicate (King 2004, Bollen, Sompel et al. 2009, Moed 2010, Leydesdorff, Bornmann et al. 2011, Kaur, Radicchi et al. 2013). They convey extrinsic characteristics of research.

In contrast, scholars have examined prominent scientific discoveries in great detail from historical, sociological, and philosophical viewpoints. Studies in this category aim to reveal intrinsic patterns that convey insights into critical paths leading to a breakthrough (Kuhn 1962) or foresights into future developments (Martin 2010). We will not be able to appreciate the significance of scholarly work until we learn about the perspective of the scholar, the focus of the attention, and the context of its origin.

A profound challenge to integrate the indicative power of research metrics and the insight-seeking analytic approaches is the difficulty in linking two perspectives that differ in so many ways at so many levels. A single perspective is not capable of characterizing and conveying the breadth and the depth of scholarly activities. Aggregation is often necessary but important details may be lost.

A problem of great challenge in one perspective may become resolvable in another. Field normalization, for example, has been intensively studied for improving the universality of research metrics. Drawing the boundary of a field or a discipline is notoriously hard. A more effective method may require a holistic view of interconnected disciplines. Many research questions may benefit from reconciling seemingly contradictory information. Until we are able to move back and forth between distinct perspectives efficiently and effectively, our ability to fully utilize the value of the scholarly knowledge that so many have spent so much effort to obtain would be rather limited.

In summary, the challenges outlined above illustrate the diverse range of theoretical and practical questions that may stimulate not only the study of scholarly metrics and analytics, but also the practice of research assessment, science policy, and many other aspects of our society. There are many more challenges ahead. Setting the study of scholarly metrics and analytics on a holistic and integrative stage is a step towards fostering creative and impactful interactions between distinct perspectives and viewpoints.